\documentclass[10pt, twocolumn, nofootinbib,superscriptaddress]{revtex4-1}
\usepackage{amsmath,amssymb,amsfonts}
\usepackage{algorithmic}
\usepackage{graphicx}
\usepackage{textcomp}
\usepackage{xcolor}
\usepackage{ragged2e}
\usepackage{booktabs, makecell, tabularx}

\usepackage{gensymb}
\makeatletter
    \renewcommand\@make@capt@title[2]{%
     \@ifx@empty\float@link{\@firstofone}{\expandafter\href\expandafter{\float@link}}%
      {\textbf{#1}}\@caption@fignum@sep#2\quad}%
\makeatother
\makeatletter 
\renewcommand{\fnum@figure}{\textbf{Fig.~\thefigure}}
\makeatother
\usepackage{xcolor}
\newcommand{\beginsupplement}{%
        \setcounter{table}{0}
        \renewcommand{\thetable}{S\arabic{table}}%
        \setcounter{figure}{0}
        \renewcommand{\thefigure}{S\arabic{figure}}%
        \setcounter{equation}{0}
        \renewcommand{\theequation}{S\arabic{equation}}%
     }
\def\BibTeX{{\rm B\kern-.05em{\sc i\kern-.025em b}\kern-.08em
    T\kern-.1667em\lower.7ex\hbox{E}\kern-.125emX}}
\begin{document}

\title{Brillouin and Kerr nonlinearities of a low-index silicon oxynitride platform}

\author{Kaixuan~Ye}
\author{Yvan~Klaver}
\affiliation{Nonlinear Nanophotonics, MESA+ Institute of Nanotechnology, University of Twente, Enschede, the Netherlands}
\author{Oscar~A~Jimenez~Gordillo}
\affiliation{{Dipartimento di Elettronica, Informazione e Bioingegneria (DEIB), Politecnico di Milano, 20133, Italy}}
\author{Roel~Botter}
\author{Okky~Daulay}
\affiliation{Nonlinear Nanophotonics, MESA+ Institute of Nanotechnology, University of Twente, Enschede, the Netherlands}
\author{Francesco Morichetti}
\author{Andrea Melloni}
\affiliation{{Dipartimento di Elettronica, Informazione e Bioingegneria (DEIB), Politecnico di Milano, 20133, Italy}}
\author{David~Marpaung}
\email{Corresponding author: david.marpaung@utwente.nl}
\affiliation{Nonlinear Nanophotonics, MESA+ Institute of Nanotechnology, University of Twente, Enschede, the Netherlands}

\date{\today}

\begin{abstract}
Nonlinear optical effects including stimulated Brillouin scattering (SBS) and four-wave mixing (FWM) play an important role in microwave photonics, optical frequency combs, and quantum photonics. Harnessing SBS and FWM in a low-loss and versatile integrated platform would open the path to building large-scale Brillouin/Kerr-based photonic integrated circuits. In this letter, we investigate the Brillouin and Kerr properties of a low-index (n=1.513~@~1550~nm) silicon oxynitride (SiON) platform. We observed, for the first time, backward SBS in SiON waveguides with a Brillouin gain coefficient of 0.3~m$^{-1}$W$^{-1}$, which can potentially be increased to 0.95~m$^{-1}$W$^{-1}$  by just tailoring the waveguide cross-section. We also performed FWM experiments in SiON rings and obtained the nonlinear parameter $\gamma$, of 0.02~m$^{-1}$W$^{-1}$. Our results point to a low-loss and low-index photonic integrated platform that is both Brillouin and Kerr active.
\end{abstract}

\maketitle

\section*{Introduction}
Stimulated Brillouin scattering (SBS), which is an interaction between optical and acoustic waves, is currently revolutionizing  photonic integrated circuit designs \cite{EggletonNP2019,PantOE2011,KittlausNP2018,KittlausNC2017,KittlausNP2016,RakichPRX2012,GundavarapuNP2018,BotterSciAdv2022}. Featuring a narrow-band (tens of MHz) gain resonance shifted around tens of GHz away from the pump light, the on-chip SBS plays a significant role in microwave photonics \cite{MarpaungNP2019,Marpaung2015,McKayOptica2019}, narrow-linewidth integrated lasers \cite{OtterstromScience2018,GundavarapuNP2018,ChauhanNC2021}, and on-chip nonreciprocal light propagation \cite{KittlausNP2018,KimNP2015}.

Efficient on-chip SBS process requires simultaneously guiding both the optical and gigahertz acoustic waves in a waveguide, making it challenging to be realized in most integrated platforms. 
Several encouraging results have been demonstrated recently in various platforms, including chalcogenide \cite{PantOE2011}, silicon \cite{KittlausNP2016}, doped silica \cite{LiPTL2020}, aluminum gallium arsenide \cite{JinCLEO2020}, and aluminum nitride \cite{LiuOptica2019}. In addition, SBS has also been observed in silicon nitride-based waveguides \cite{GundavarapuNP2018,GygerPRL2020,BotterSciAdv2022}, opening the pathway to intersect Brillouin scattering with Kerr nonlinearities in low-loss and mature  platforms. 

\begin{figure}[t!]
\centering
\includegraphics{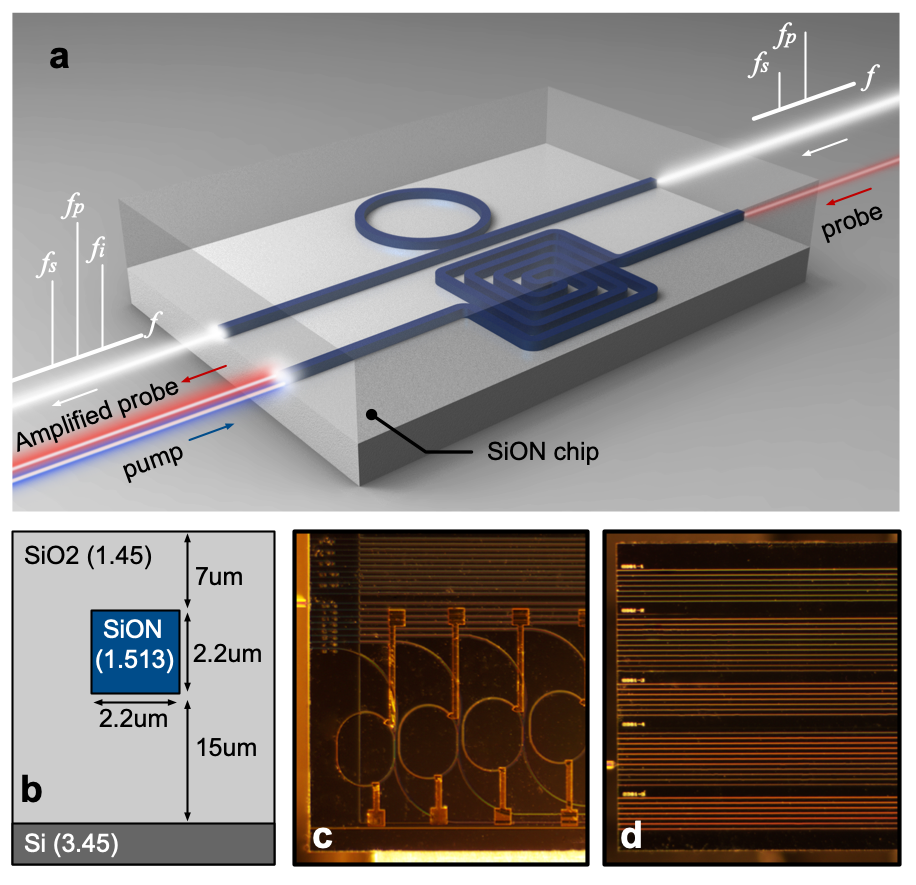}
\caption{(a) Artistic representation of the SiON waveguides, showing the four-wave mixing process in an all-pass microring resonator and the backward stimulated Brillouin scattering (SBS) in a spiral waveguide. (b) The cross-section of the SiON platform in our work. (c) The chip photograph of the SiON microring resonators with a FSR of 50 GHz. (d) The chip photograph of the 5-cm SiON straight waveguide.}
\label{fig1}
\end{figure}

\begin{figure}[t!]
\centering
\includegraphics[width=\linewidth]{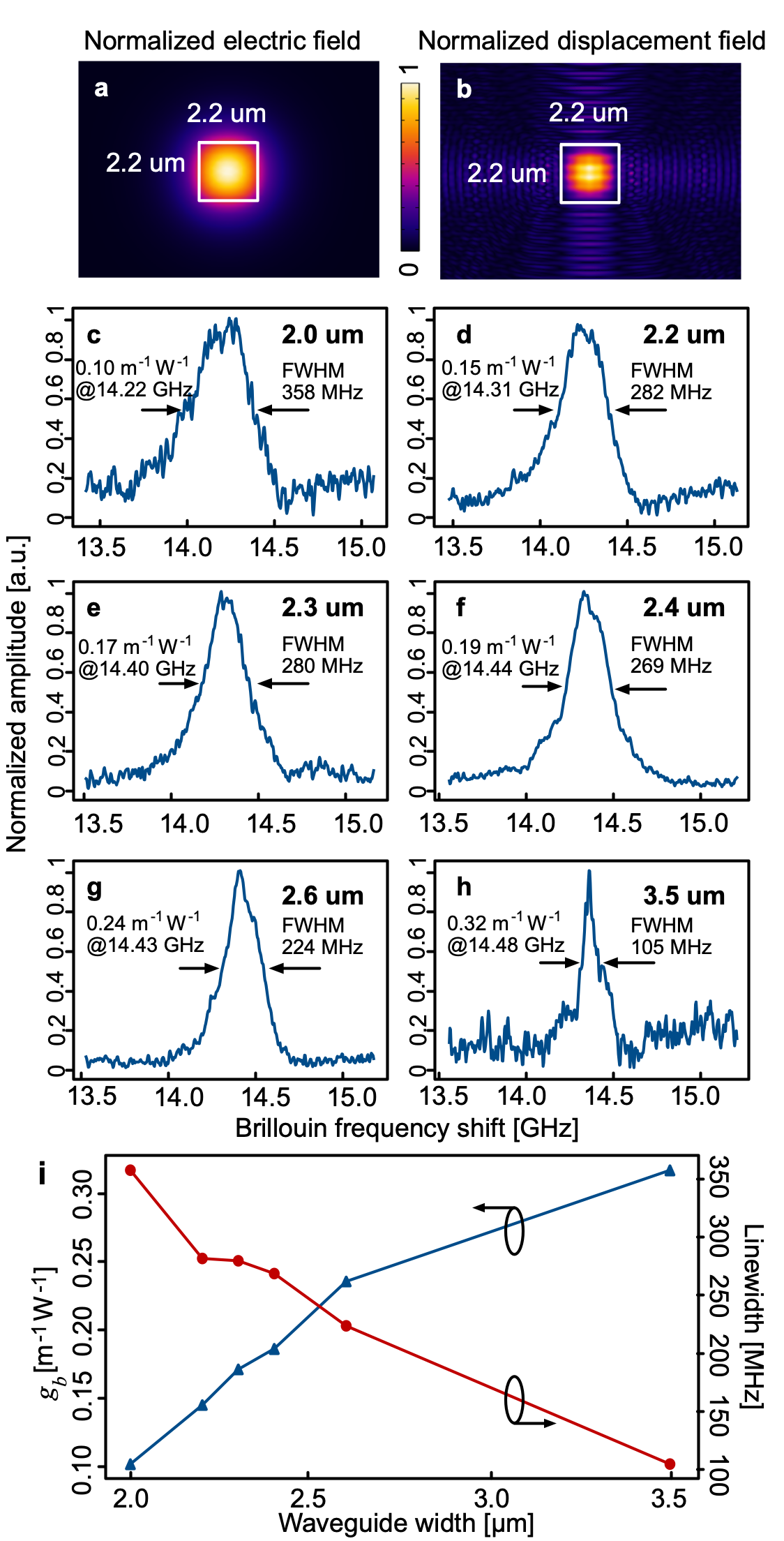}
\caption{(a) Simulated optical mode of the SiON waveguide. (b) Simulated acoustic response of the SiON waveguide. (c)-(h) Measured SBS gain spectra of the 2.0~µm, 2.2~µm, 2.3~µm, 2.4~µm, 2.6~µm, and 3.5~µm-wide SiON waveguides, respectively. (i) Brillouin gain coefficients and linewidth of the SiON waveguides with different widths.}
\label{fig2}
\end{figure}

Silicon oxynitride (SiON) is another highly-developed integrated platform that has appealing properties including low propagation loss, wide transparency window, absence of multi-photon absorption effects, and stress-free fabrication \cite{MossNP2013,TrentiAIP2018}. 

The optical and mechanical properties of SiON could be tuned continuously between those of SiO$_2$ and Si$_3$N$_4$ at different nitrogen/oxygen (N/O) ratios \cite{BaakAO1982,GrahnAPL1998}. For example, a variety of SiON, known as Hydex (n=1.7~@~1550~nm), has been widely used for Kerr-based nonlinear optic applications including optical frequency comb \cite{RowleyNature2022}, optical neural network \cite{XuNature2021}, and quantum photonics \cite{ReimerScience2016}. A slightly higher index SiON (n=1.83~@~1550~nm) was also proposed in \cite{TrentiAIP2018,PiccoliOME2022} for Kerr-based applications. In both cases, the SiON platforms have a refractive index close to silicon nitride (n=1.98~@~1550~nm) instead of silicon oxide (n=1.45~@~1550nm). The relatively high refractive index induces a high nonlinear index, making it useful for Kerr-based nonlinear optic applications.

But from the Brillouin perspectives, high refractive index SiON is less attractive due to  the high content of the nitrogen that leads to a meager photoelastic coefficient p$_{12}$ because of the weak p$_{12}$ of the Si$_3$N$_4$ \cite{GygerPRL2020}. Moreover, high-index SiON also has similar mechanical properties to Si$_3$N$_4$, such as high acoustic velocity that prevents acoustic confinement when cladded with SiO$_2$  \cite{GundavarapuNP2018,GygerPRL2020,BotterSciAdv2022}.

In this paper, we investigate the Brillouin and Kerr properties of a SiON integrated platform with a relatively lower refractive index (n=1.513~@~1550~nm). Contrasting to SiON platforms mentioned above, the SiON platform investigated here has a larger photoelastic coefficient p$_{12}$, lower acoustic velocity, and a larger cross-section, all of which lead to an enhanced SBS effect. We experimentally observed, for the first time to our knowledge, backward SBS in SiON waveguides. We also characterized the Brillouin gain coefficient $g_b$ of the SiON waveguides with different widths. We found out the $g_b$ of this SiON waveguide can potentially be increased to around 0.95~m$^{-1}$W$^{-1}$ by simply tailoring the waveguide cross-section. This sufficiently large Brillouin gain coefficient, together with the low propagation loss, makes it possible to generate decent SBS gain for a plethora of Brillouin-based applications in this SiON platform. 

Furthermore, we also measured the nonlinear parameter $\gamma$ and nonlinear index $n_2$ of this SiON platform through four-wave mixing (FWM) experiments in a ring resonator. While the measured $\gamma$ is an order of magnitude lower when compared to that of high-index SiON, we expect that with lower losses and higher pump power, the unique interplay between the SBS and Kerr effect such as  Brillouin-assisted Kerr frequency comb \cite{Bai2021, Nie2022}  could be observed in this integrated platform.

\section*{Results}
We performed the backward SBS and four-wave mixing experiments in single-pass (spiral or straight) waveguides and microring resonators respectively, as shown in Fig.~\ref{fig1}(a). The cross-section of this platform is shown in Fig.~\ref{fig1}(b) \cite{MorichettiACS2016,Morichetti2007}. The 2.2~µm-thick SiON layer has a refractive index $n$ of 1.513 at 1550~nm. It is on top of a 15-µm SiO$_2$ layer and is covered by a 7~µm-thick SiO$_2$ upper cladding. The refractive index contrast $\Delta n$ between the core and the cladding is 4.4\%, enabling a bending radius of 600~µm with negligible radiation losses. Fig.~\ref{fig1}(c) shows the photograph of the microring resonators in this platform with a free spectral range (FSR) of 50 GHz and coupling coefficients varying from 0.05 to 0.8. Fig.~\ref{fig1}(d) shows the photograph of several groups of 5-cm straight waveguides with different widths. The measured propagation loss of those straight waveguides is 0.25~dB/cm with coupling loss to lensed-tip fibers of approximately 3~dB/facet.

\subsection*{Stimulated Brillouin Scattering in SiON Waveguides}\label{sec:SBS}
We developed a finite element model \cite{BotterSciAdv2022} in COMSOL to estimate the SBS response of the SiON waveguides. The simulated optical field and the corresponding acoustic response of the 2.2~µm-wide SiON waveguide are shown in Fig.~\ref{fig2}(a) and (b), respectively. The optical field is well confined around the SiON core area because of the total internal reflection (TIR). However, the TIR condition does not hold for the acoustic response because the acoustic velocity of the SiON ($\sim$ 6.2~km/s) is higher than that of the SiO$_2$ ($\sim$ 5.9~km/s). As a result, part of the acoustic field would leak into the cladding as shown in Fig.~\ref{fig2}(b). Nevertheless, most of the acoustic field still remains inside the SiON core because of the relatively large cross-section area \cite{PoultonJOSAB2013}. This results in a large overlap between the optical and acoustic fields that leads to improved Brillouin gain coefficient. Extensive simulation results of the SBS gain coefficients are included in the Supplementary. 

To verify the simulation results, we characterized the SBS responses of the SiON waveguides with a pump-probe experimental apparatus \cite{BotterSciAdv2022,GygerPRL2020}. The pump and probe light are intensity-modulated and coupled into the opposite facets of the waveguide. We keep the pump frequency fixed at 1561 nm while sweeping the probe at frequencies down shifted from the pump by about 15~GHz. When the frequency difference between the pump and the probe is close to the Brillouin frequency shift of the SiON waveguide, the probe will experience the SBS gain and a peak will be detected at the lock-in amplifier (See the Supplementary for more details about the SBS experiment). 

\begin{figure}[t!]
\centering
\includegraphics[width=\linewidth]{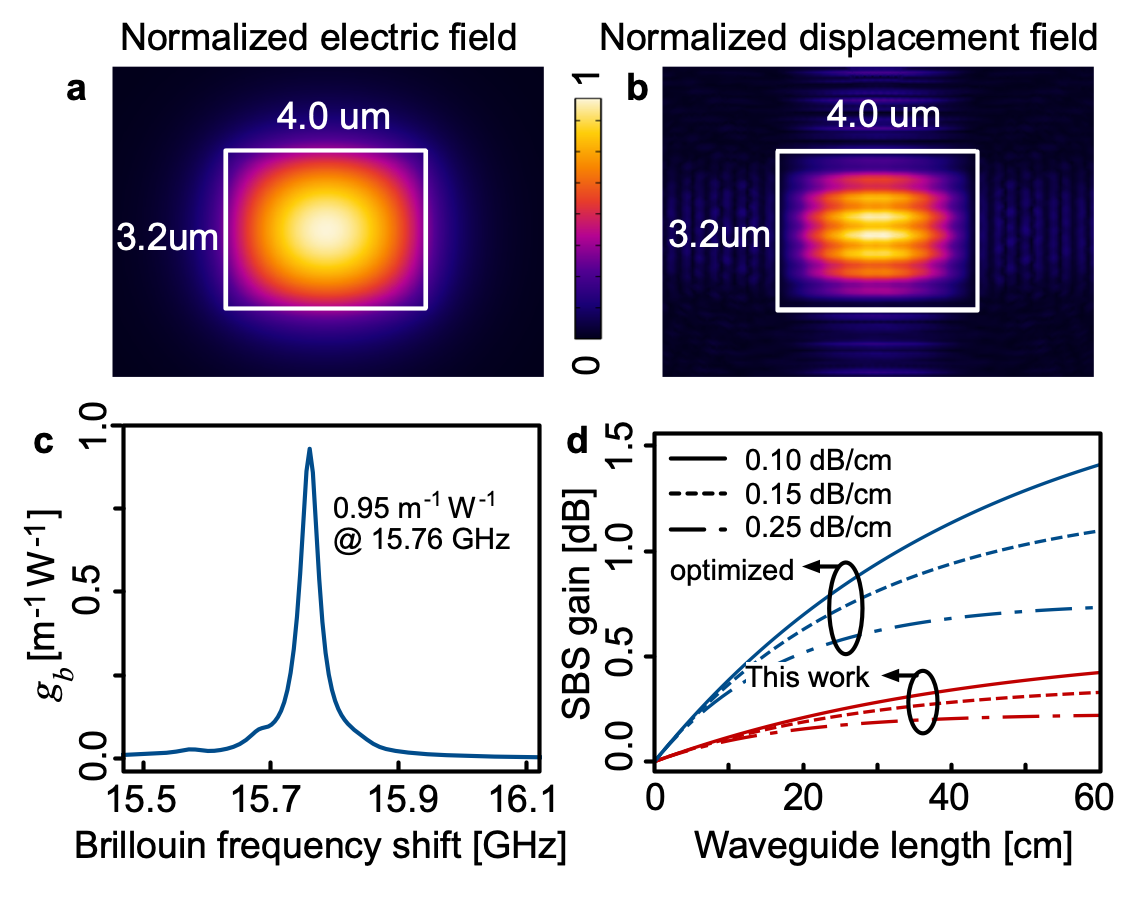}
\caption{(a) Simulated optical mode and (b) simulated acoustic response and (c) simulated Brillouin gain spectrum of the optimized SiON waveguide. (d) Estimated SBS gain from the optimized and current SiON waveguides.}
\label{fig3}
\end{figure}

Several 5~cm-long SiON waveguides are characterized to investigate the influence of waveguide width on the Brillouin gain spectra.  The measured SBS responses of the 2.0~µm, 2.2~µm, 2.3~µm, 2.4~µm, 2.6~µm, and 3.5~µm-wide waveguides are shown in Fig.~\ref{fig2}(c) to (h), respectively. All waveguides show a clear SBS peak well above the noise floor with the Brillouin frequency shift increases from 14.22 GHz for the 2.0~µm-wide waveguide to 14.48~GHz for the 3.5~µm-wide waveguide. Fig.~\ref{fig2}(i) plots the measured Brillouin gain coefficient $g_b$ and the SBS linewidth of the SiON waveguides with different widths (See the Supplementary for more details about the Brillouin gain coefficient calculation). The Brillouin gain coefficient $g_b$ increases from 0.1~m$^{-1}$W$^{-1}$ to 0.32~m$^{-1}$W$^{-1}$ when the waveguide width increases from 2.0~µm to 3.5~µm. In the meantime, the linewidth of the SBS peak reduces from 358~MHz to 105~MHz. The increasing Brillouin gain coefficient and the narrowing of the SBS linewidth indicate an improvement in acoustic confinement when the SiON waveguides become wider. 

The Brillouin gain coefficient can be further increased by optimizing the cross-section of the waveguide through the genetic algorithm \cite{BotterSciAdv2022}. Fig.~\ref{fig3}~(a) and (b) show the simulated optical mode and the acoustic response of a SiON waveguide with the same core refractive index but with an optimized cross-section for SBS gain. The dimension of such a waveguide is 4.0~µm $\times$ 3.2~µm with a top cladding of 3~µm and a bottom cladding of 10~µm. Compared to the optical and acoustic fields of the waveguide structure in this work, less acoustic field is scattered into the cladding while the optical field is still well confined in the optimized waveguide structure. The Brillouin gain spectrum of the optimized waveguide structure is shown in Fig.~\ref{fig3}~(c). The simulated peak Brillouin gain coefficient of this waveguide is 0.95~m$^{-1}$W$^{-1}$, which is 3$\times$ higher than the waveguide structure measured in this work. Furthermore, the propagation loss in this SiON platform can also be significantly lowered by reducing sidewall roughness and improving the thermal annealing process \cite{Morichetti2007}, allowing for longer effective waveguide length for the SBS process. Fig.~\ref{fig3}~(d) estimates the SBS gain of both the measured and the optimized SiON waveguides with different propagation losses. The optimized Brillouin gain coefficient (around 0.95~m$^{-1}$W$^{-1}$), along with the improved propagation loss (around 0.1~dB/cm), can enhance the SBS gain from less than 0.5~dB to near 1.5~dB for a 60-cm waveguide, which is sufficient for applications like SBS-based narrow-bandwidth microwave photonic notch filters \cite{BotterSciAdv2022,Marpaung2015}.


\begin{figure}[t!]
\centering
\includegraphics[width=\linewidth]{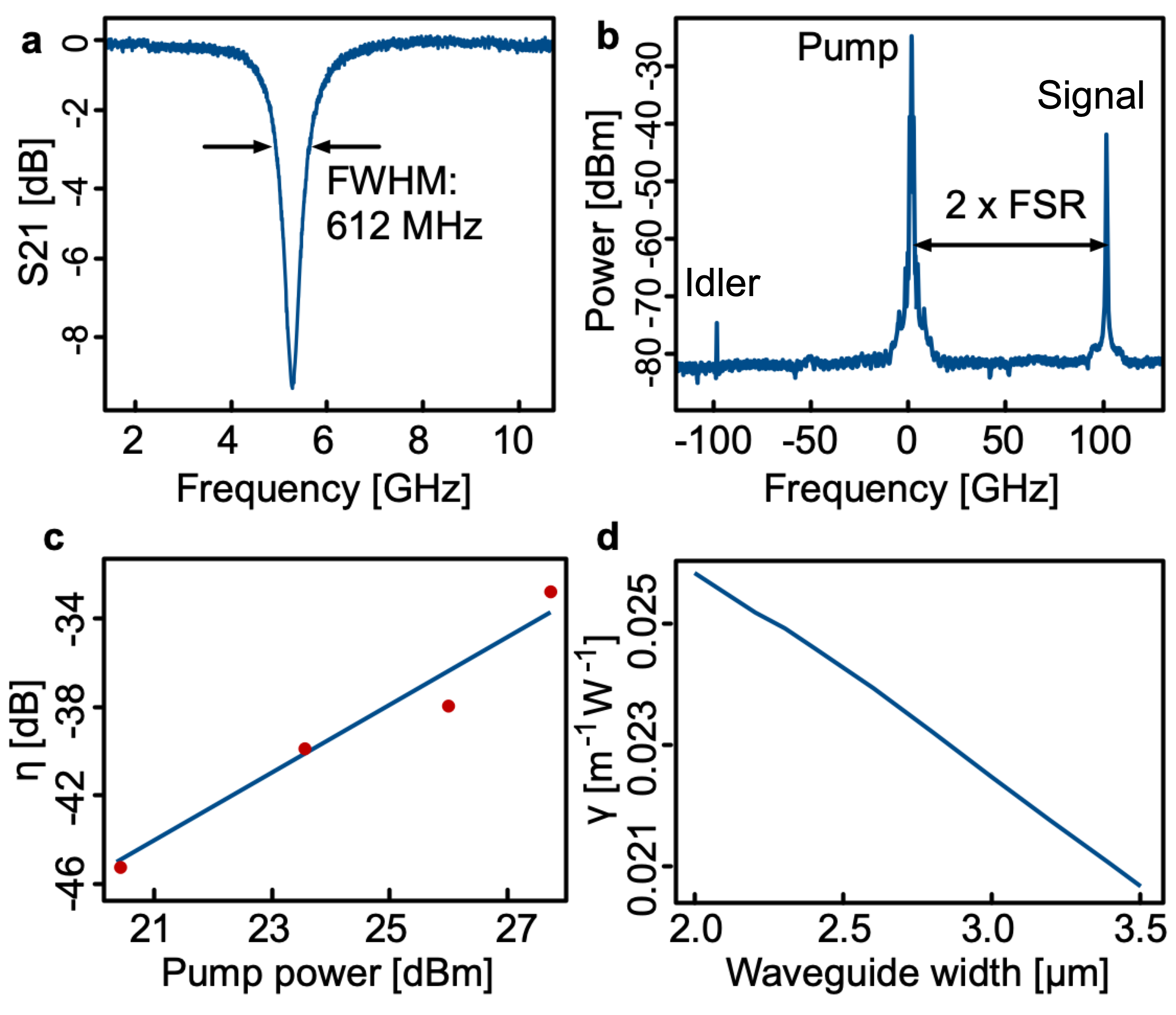}
\caption{(a) Measured resonance response of the SiON ring resonator. (b) Measured four-wave mixing response of the SiON ring resonator. (c) Conversion efficiency of the four-wave mixing at different pump power. (d) The estimated nonlinear parameter $\gamma$ of the SiON waveguides with different widths.}
\label{fig4}
\end{figure}

\subsection*{Four-wave mixing in SiON Waveguides}\label{sec:FWM}
We further investigate the Kerr nonlinearities of this SiON platform. High-index SiON platforms are widely used for Kerr-based nonlinear optics applications because of the relatively large nonlinear parameter $\gamma$ \cite{MossNP2013}. However, the nonlinear parameter $\gamma$ is highly dependent on the refractive index  and the geometry of the waveguide. The SiON waveguide in this work has a relatively lower refractive index and a larger cross-section compared with other SiON platforms \cite{MossNP2013,TrentiAIP2018}, and the nonlinear index $n_2$ and nonlinear parameter $\gamma$ of the SiON waveguide in this platform has never been characterized before. 

We devised a four-wave mixing (FWM) experiment for the nonlinear parameter characterization. Because of the limited effective length of the available samples, the FWM conversion efficiency of the straight waveguide is comparable with that of the fiber pigtails, making it difficult to accurately measure the $n_2$ and the $\gamma$. We use the all-pass ring resonators to enhance the FWM in the SiON waveguide so that the contribution from fibers in the setup can be neglected \cite{Absil2000}. The ring resonator applied in our experiment is made of the 2.2~µm-wide SiON waveguide and it has a free spectral range (FSR) of 50~GHz and a power coupling coefficient of 0.05. The pump laser is locked close to the resonance of the ring resonator to mitigate the thermal influence on the ring resonator. The signal laser is set close to 2~$\times$ FSR away from the pump signal and is combined with the pump light with a 99:1 coupler. The combined pump and signal are coupled into the all-pass ring resonator with a lensed fiber with a spot size of 2~µm. The generated idler is then coupled out from the chip and sent to the optical spectrum analyzer to measure the conversion efficiency from the signal to the generated idler (See the Supplementary for details of the FWM experiment).

To determine the field enhancement factor of the FWM process in the ring resonator, we first characterized the resonance response of the ring resonator with a vector network analyzer, as shown in Fig.~\ref{fig4}~(a) (See the Supplementary for details of the characterization). The measured full-width at half-maximum (FWHM) is 612~MHz with an extinction ratio of 8.9~dB, corresponding to a loaded Q-factor of 330,000 and a propagation loss of 0.27~dB/cm. Fig.~\ref{fig4}~(b) shows the measured FWM response of the 50~GHz SiON ring resonator. A clear peak is shown at 2 $\times$ FSR down shifted from the pump frequency, which is the idler generated from the FWM process between the pump and signal in the ring resonator. 

The nonlinear index $n_2$ and nonlinear parameter $\gamma$ of the SiON waveguide in this platform can be estimated from the conversion efficiency between the signal and the idler (See the supplementary for details of the calculation). Fig.~\ref{fig4}~(c) shows the measured conversion efficiency of the FWM process at different pump power. Based on this measurement, the calculated $\gamma$ and $n_2$ of the 2.2~µm-wide SiON waveguide are 0.024~m$^{-1}$W$^{-1}$ and 4.16~$\times 10^{-20}$~m$^{2}$/W, respectively. We also estimated the nonlinear parameter $\gamma$ of the SiON waveguides with different widths based on the measured value of $n_2$, as shown in Fig.~\ref{fig4}~(d). The $\gamma$ decreases from around 0.025~m$^{-1}$W$^{-1}$ to 0.020~m$^{-1}$W$^{-1}$ when the waveguide width reduces from 2.0~µm to 3.5~µm.

\section*{Discussion}
For Brillouin-Kerr interactions, the balance between the nonlinearities needs to be considered. In microcavities, it is generally preferred to have larger Brillouin gain, as it is easier to inhibit cascading or other unwanted interactions via mode manipulation. Comparing the values of the measured  $g_b$ in Fig.~\ref{fig2}~(i) and $\gamma$ in Fig.~\ref{fig4}~(a), the SiON waveguides reported here have an order of magnitude larger Brillouin gain compared to Kerr nonlinearity. This $g_b/\gamma$ ratio is similar to previous demonstrations of Brillouin-assisted Kerr frequency combs in \cite{Bai2021, Nie2022}, showing the potential to realize it in an integrated platform.


In conclusion, we have investigated the Brillouin and Kerr properties of a SiON integrated platform with a relatively low refractive index. We observed, for the first time, the backward SBS response of those SiON waveguides. We also measured its nonlinear index $n_2$ and nonlinear parameter $\gamma$. These SiON waveguides can be fabricated in a versatile and low-loss integrated platform, and can potentially lead to a plethora of Brillouin and Kerr-based applications, including narrow-bandwidth microwave photonic filters, and narrow-linewidth lasers, and optical frequency combs.

\section*{Author Contributions}
D.M. and K.Y. developed the concept and proposed the physical system. K.Y. and Y.K. developed and performed numerical simulations. K.Y. performed the SBS characterisation with input from R.B., K.Y., and O.D. Y.K. and K.Y. performed the FWM experiments. O.A.J.G., F.M., and A.M. developed and fabricated the samples. K.Y., D.M., and Y.K. wrote the manuscript. D.M. led and supervised the entire project.

\section*{Funding Information}
This project is funded by the European Research Council Consolidator Grant (101043229 TRIFFIC) and Nederlandse Organisatie voor Wetenschappelijk Onderzoek (NWO) projects (740.018.021 and 15702).

\bibliographystyle{IEEEtran} 
\bibliography{library}

\begin{thebibliography}{10}
\providecommand{\url}[1]{#1}
\csname url@samestyle\endcsname
\providecommand{\newblock}{\relax}
\providecommand{\bibinfo}[2]{#2}
\providecommand{\BIBentrySTDinterwordspacing}{\spaceskip=0pt\relax}
\providecommand{\BIBentryALTinterwordstretchfactor}{4}
\providecommand{\BIBentryALTinterwordspacing}{\spaceskip=\fontdimen2\font plus
\BIBentryALTinterwordstretchfactor\fontdimen3\font minus
  \fontdimen4\font\relax}
\providecommand{\BIBforeignlanguage}[2]{{%
\expandafter\ifx\csname l@#1\endcsname\relax
\typeout{** WARNING: IEEEtran.bst: No hyphenation pattern has been}%
\typeout{** loaded for the language `#1'. Using the pattern for}%
\typeout{** the default language instead.}%
\else
\language=\csname l@#1\endcsname
\fi
#2}}
\providecommand{\BIBdecl}{\relax}
\BIBdecl

\bibitem{EggletonNP2019}
B.~J. Eggleton, C.~G. Poulton, P.~T. Rakich, M.~J. Steel, and G.~Bahl,
  ``Brillouin integrated photonics,'' \emph{Nature Photonics}, 2019.

\bibitem{PantOE2011}
R.~Pant, C.~G. Poulton, D.-Y. Choi \emph{et~al.}, ``On-chip stimulated
  brillouin scattering,'' \emph{Optics Express}, vol.~19, pp. 8285--8290, 4
  2011.

\bibitem{KittlausNP2018}
E.~A. Kittlaus, N.~T. Otterstrom, P.~Kharel, S.~Gertler, and P.~T. Rakich,
  ``Non-reciprocal interband brillouin modulation,'' \emph{Nature Photonics},
  vol.~12, pp. 613--619, 2018.

\bibitem{KittlausNC2017}
E.~A. Kittlaus, N.~T. Otterstrom, and P.~T. Rakich, ``On-chip inter-modal
  brillouin scattering,'' \emph{Nature Communications}, vol.~8, pp. 1--9, 2017.

\bibitem{KittlausNP2016}
E.~A. Kittlaus, H.~Shin, and P.~T. Rakich, ``Large brillouin amplification in
  silicon,'' \emph{Nature Photonics}, vol.~10, pp. 463--467, 2016.

\bibitem{RakichPRX2012}
P.~T. Rakich, C.~Reinke, R.~Camacho, P.~Davids, and Z.~Wang, ``Giant
  enhancement of stimulated brillouin scattering in the subwavelength limit,''
  \emph{Physical Review X}, vol.~2, p. 11008, 2012.

\bibitem{GundavarapuNP2018}
S.~Gundavarapu, G.~M. Brodnik, M.~Puckett \emph{et~al.}, ``Sub-hertz
  fundamental linewidth photonic integrated brillouin laser,'' \emph{Nature
  Photonics}, vol.~13, pp. 60--67, 12 2018.

\bibitem{BotterSciAdv2022}
R.~Botter, K.~Ye, Y.~Klaver \emph{et~al.}, ``Guided-acoustic stimulated
  brillouin scattering in silicon nitride photonic circuits,'' \emph{Science
  Advances}, vol.~8, p. 2196, 10 2022.

\bibitem{MarpaungNP2019}
D.~Marpaung, J.~Yao, and J.~Capmany, ``Integrated microwave photonics,''
  \emph{Nature Photonics}, vol.~13, pp. 80--90, 2019.

\bibitem{Marpaung2015}
D.~Marpaung, B.~Morrison, M.~Pagani \emph{et~al.}, ``Low-power, chip-based
  stimulated brillouin scattering microwave photonic filter with ultrahigh
  selectivity,'' \emph{Optica}, vol.~2, p.~76, 2015.

\bibitem{McKayOptica2019}
L.~McKay, M.~Merklein, A.~C. Bedoya \emph{et~al.}, ``Brillouin-based phase
  shifter in a silicon waveguide,'' \emph{Optica, Vol. 6, Issue 7, pp.
  907-913}, vol.~6, pp. 907--913, 7 2019.

\bibitem{OtterstromScience2018}
N.~T. Otterstrom, R.~O. Behunin, E.~A. Kittlaus, Z.~Wang, and P.~T. Rakich, ``A
  silicon brillouin laser,'' \emph{Science}, vol. 360, pp. 1113--1116, 6 2018.

\bibitem{ChauhanNC2021}
N.~Chauhan, A.~Isichenko, K.~Liu \emph{et~al.}, ``Visible light photonic
  integrated brillouin laser,'' \emph{Nature Communications 2021 12:1},
  vol.~12, pp. 1--8, 8 2021.

\bibitem{KimNP2015}
J.~Kim, M.~C. Kuzyk, K.~Han, H.~Wang, and G.~Bahl, ``Non-reciprocal brillouin
  scattering induced transparency,'' \emph{Nature Physics}, vol.~11, pp.
  275--280, 1 2015.

\bibitem{LiPTL2020}
S.~Li, X.~Li, W.~Zhang, J.~Chen, and W.~Zou, ``Investigation of brillouin
  properties in high-loss doped silica waveguides by comparison experiment,''
  \emph{IEEE Photonics Technology Letters}, vol.~32, pp. 948--951, 2020.

\bibitem{JinCLEO2020}
W.~Jin, L.~Chang, W.~Xie \emph{et~al.}, ``Stimulated brillouin scattering in
  algaas on insulator waveguides,'' \emph{Conference on Lasers and
  Electro-Optics, paper SM4L.7}, 2020.

\bibitem{LiuOptica2019}
Q.~Liu, H.~Li, and M.~Li, ``Electromechanical brillouin scattering in
  integrated optomechanical waveguides,'' \emph{Optica}, vol.~6, pp. 778--785,
  2019.

\bibitem{GygerPRL2020}
F.~Gyger, J.~Liu, F.~Yang \emph{et~al.}, ``Observation of stimulated brillouin
  scattering in silicon nitride integrated waveguides,'' \emph{Physical Review
  Letters}, vol. 124, 2020.

\bibitem{MossNP2013}
D.~J. Moss, R.~Morandotti, A.~L. Gaeta, and M.~Lipson, ``New cmos-compatible
  platforms based on silicon nitride and hydex for nonlinear optics,''
  \emph{Nature Photonics}, vol.~7, pp. 597--607, 7 2013.

\bibitem{TrentiAIP2018}
A.~Trenti, M.~Borghi, S.~Biasi \emph{et~al.}, ``Thermo-optic coefficient and
  nonlinear refractive index of silicon oxynitride waveguides,'' \emph{AIP
  Advances}, vol.~8, p. 025311, 2 2018.

\bibitem{BaakAO1982}
T.~Baak, ``Silicon oxynitride; a material for grin optics,'' \emph{Applied
  Optics, Vol. 21, Issue 6, pp. 1069-1072}, vol.~21, pp. 1069--1072, 3 1982.

\bibitem{GrahnAPL1998}
H.~T. Grahn, H.~J. Maris, J.~Tauc, and K.~S. Hatton, ``Elastic properties of
  silicon oxynitride films determined by picosecond acoustics,'' \emph{Applied
  Physics Letters}, vol.~53, p. 2281, 12 1998.

\bibitem{RowleyNature2022}
M.~Rowley, P.~H. Hanzard, A.~Cutrona \emph{et~al.}, ``Self-emergence of robust
  solitons in a microcavity,'' \emph{Nature 2022 608:7922}, vol. 608, pp.
  303--309, 8 2022.

\bibitem{XuNature2021}
X.~Xu, M.~Tan, B.~Corcoran \emph{et~al.}, ``11 tops photonic convolutional
  accelerator for optical neural networks,'' \emph{Nature 2020 589:7840}, vol.
  589, pp. 44--51, 1 2021.

\bibitem{ReimerScience2016}
C.~Reimer, M.~Kues, P.~Roztocki \emph{et~al.}, ``Generation of multiphoton
  entangled quantum states by means of integrated frequency combs,''
  \emph{Science}, vol. 351, pp. 1176--1180, 3 2016.

\bibitem{PiccoliOME2022}
G.~Piccoli, M.~Sanna, M.~Borghi, L.~Pavesi, and M.~Ghulinyan, ``Silicon
  oxynitride platform for linear and nonlinear photonics at nir wavelengths,''
  \emph{Optical Materials Express}, vol.~12, 2022.

\bibitem{Bai2021}
Y.~Bai, M.~Zhang, Q.~Shi \emph{et~al.}, ``Brillouin-kerr soliton frequency
  combs in an optical microresonator,'' \emph{Phys. Rev. Lett.}, vol. 126, p.
  063901, Feb 2021.

\bibitem{Nie2022}
M.~Nie, K.~Jia, Y.~Xie \emph{et~al.}, ``Synthesized spatiotemporal mode-locking
  and photonic flywheel in multimode mesoresonators,'' \emph{Nature
  Communications 2022 13:1}, vol.~13, pp. 1--9, 10 2022.

\bibitem{MorichettiACS2016}
F.~Morichetti, S.~Grillanda, S.~Manandhar \emph{et~al.}, ``Alpha radiation
  effects on silicon oxynitride waveguides,'' \emph{ACS Photonics}, vol.~3, pp.
  1569--1574, 9 2016.

\bibitem{Morichetti2007}
F.~Morichetti, A.~Melloni, A.~Breda \emph{et~al.}, ``A reconfigurable
  architecture for continuously variable optical slow-wave delay lines,''
  \emph{Optics Express}, vol.~15, pp. 17\,273--17\,282, 12 2007.

\bibitem{PoultonJOSAB2013}
C.~G. Poulton, R.~Pant, and B.~J. Eggleton, ``Acoustic confinement and
  stimulated brillouin scattering in integrated optical waveguides,''
  \emph{JOSA B}, vol.~30, pp. 2657--2664, 10 2013.

\bibitem{Absil2000}
P.~P. Absil, J.~V. Hryniewicz, B.~E. Little \emph{et~al.}, ``Wavelength
  conversion in gaas micro-ring resonators,'' \emph{Optics Letters}, vol.~25,
  pp. 554--556, 4 2000.

\end{thebibliography}

\newpage
\onecolumngrid
\beginsupplement
\newpage

\section*{Supplementary Note A: Details of the SBS Experiments}
\subsection*{Experiment setup}

\begin{figure}[h]
\centering
\includegraphics[width=0.8\linewidth]{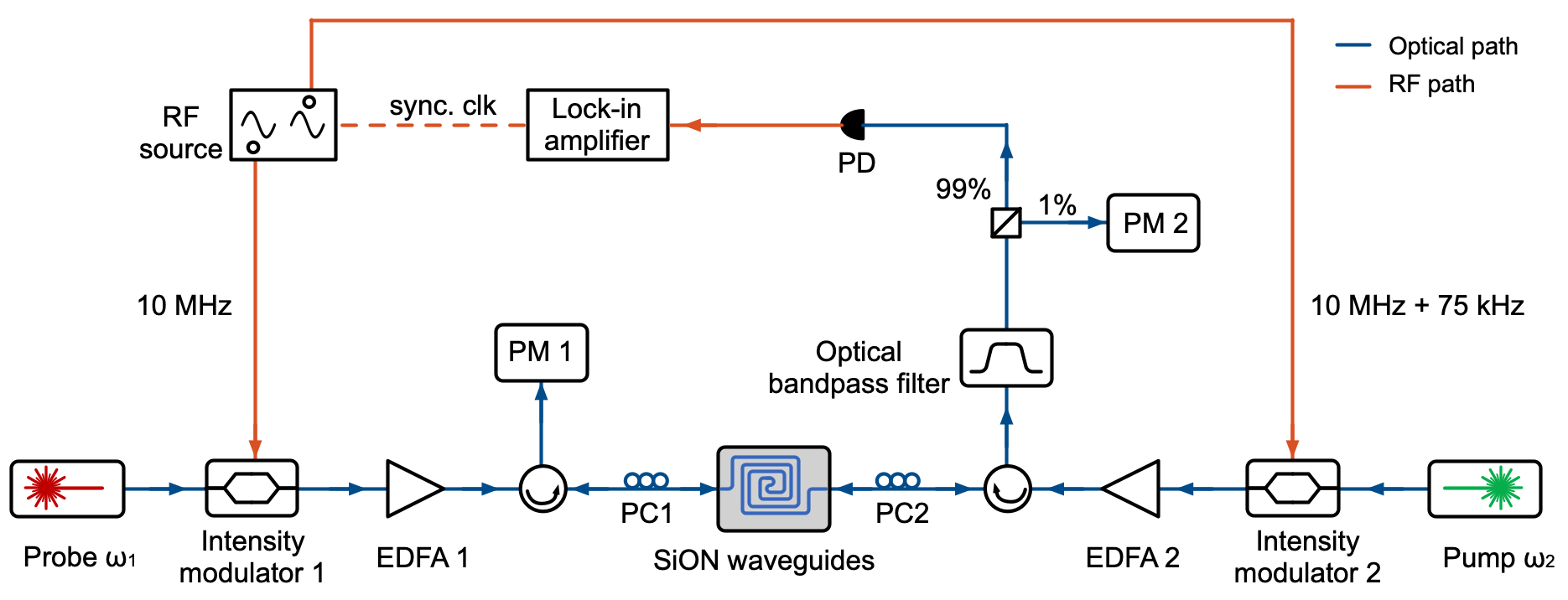}
\caption{Schematic of the setup used for the SBS characterization. EDFA: erbium-doped fiber amplifier, PM: optical power meter, PC: fiber polarization controller, PD: photodetector, RF: radiofrequency signal generator.}
\label{figs1}
\end{figure}

We applied the double-intensity-modulation pump-probe technique to characterize the Brillouin gain coefficient of the SiON 5-cm straight waveguides with different widths. Fig.~\ref{figs1} shows the schematic of the experimental setup. The pump laser (Agere D2525P22) operates at 1562~nm and is modulated by an intensity modulator (Thorlabs LN81S-FC) with a 10.075~MHz sine signal generated by an RF signal generator (Keysight EDU33212A). The pump signal is then amplified by an Erbium-doped fiber amplifier (EDFA, Amonics AEDFA-33-B-FA) to  28.7 dBm. After that, the pump signal passes an optical circulator (Thorlabs 6015-3-APC) and a polarization controller (Thorlabs FPC032) before it is coupled into the chip with an AR-coated polarization maintaining lensed fiber with a spot size of 2~$\mu$m (OZ optics). The transmitted pump power is monitored with a power meter (Thorlabs S144C). The coupling loss of the sample is 3 dB per facet.

The probe laser (Newport TLB-6728-P) sweeps at frequencies down shifted from the pump laser by about 15 GHz. The probe is modulated by an intensity modulator of the same model (Thorlabs LN81S-FC) with a 10~MHz signal generated by the other channel of the RF signal generator (Keysight EDU33212A). It is then amplified by an EDFA (Amonics AEDFA-PA-35-B-FA) to 21.6 dBm. After that, the probe signal passes an optical circulator (Thorlabs 6015-3-APC) and a polarization controller (Thorlabs FPC032) before it is coupled into the other side of the chip with an identical lensed fiber. The transmitted probe signal passes through an optical bandpass filter (EXFO XTM-50) to filter out the reflected pump. After that, 1\% of the probe signal is tapped into the power meter (Thorlabs S144C). The remaining probe signal is sent into a photodiode (Optilab PD-23-C-DC). The detected electrical signal is then analyzed with a lock-in amplifier (Zürich Instruments, MFLI 500~kHz) that is synchronized with the RF source. TABLE \ref{tab:SBS_param} lists the experimental parameters of the setup.

\begin{table}[ht]
\centering
\caption{\textbf{The experimental parameters of the SBS characterization setup.}}
\label{tab:SBS_param}
\begin{tabular}{c|c|c|c}
    \textbf{Parameter} &\textbf{Value} & \textbf{Unit} &\textbf{Description}\\
    \hline
    $P_{\rm{probe}}$ & 21.6 & dBm & Probe optical power after amplification\\
    $P_{\rm{pump}}$ & 28.7 & dBm & Pump optical power after amplification\\
    $V_{\pi,\rm{probe}}$ & 7.2 & V & $V_{\pi}$ @ DC of the probe modulator\\
    $V_{\pi,\rm{pump}}$ & 6.4 & V & $V_{\pi}$ @ DC of the pump modulator\\
    $P_{\rm{mod,probe}}$ & 0 & dBm & RF power sent to probe modulator\\
    $P_{\rm{mod,pump}}$ & 0 & dBm & RF power sent to pump modulator\\
    $r_{\rm{pd}}$ & 1.05 & A/W & Photodiode sensitivity\\
    $P_{\rm{pd}}$ & 6.0 & dBm & Optical power detected at the photodiode\\
    $\alpha_{\rm{c}}$ & 3 & dB/facet & Coupling loss per facet, including fiber components
\end{tabular}
\end{table}

\subsection*{Brillouin gain coefficient calculation}

The SBS gain in a waveguide is determined by:
\begin{equation}\label{eq:SBS}
G = e^{g_{\rm{B}}L_{\rm{eff}}P_{\rm{pump}}},
\end{equation}
where $g_{\rm{B}}$ is the Brillouin gain coefficient in m$^{-1}$W$^{-1}$, $L_{\rm{eff}}$ is the effective length of the waveguide, and $P_{\rm{pump}}$ is the on-chip pump power.

The effective length of a waveguide is calculated using:

\begin{equation}
    L_{\rm{eff}} = \frac{1-e^{-\alpha L}}{\alpha},
\end{equation}
where $\alpha$ is the propagation loss, and $L$ is the actual waveguide length.

By using \eqref{eq:SBS}, and taking the small signal approximation we can calculate the gain coefficient using:

\begin{equation}
g_{\rm{B}, \rm{SDS}} = \frac{V_{\rm{SDS}}}{V_{\rm{fiber}}}\frac{g_{\rm{B}, \rm{fiber}}L_{\rm{eff, fiber}}P_{\rm{pump,fiber}}}{L_{\rm{eff, SDS}}P_{\rm{pump,SDS}}}
\end{equation}

Here $V$ denotes the signal voltage measured by the lock-in amplifier, the subscripts \rm{fiber} and \rm{SDS} refer to the properties of the fiber and chip used in this experiment.

\begin{figure}
\centering
\includegraphics[width=0.8\linewidth]{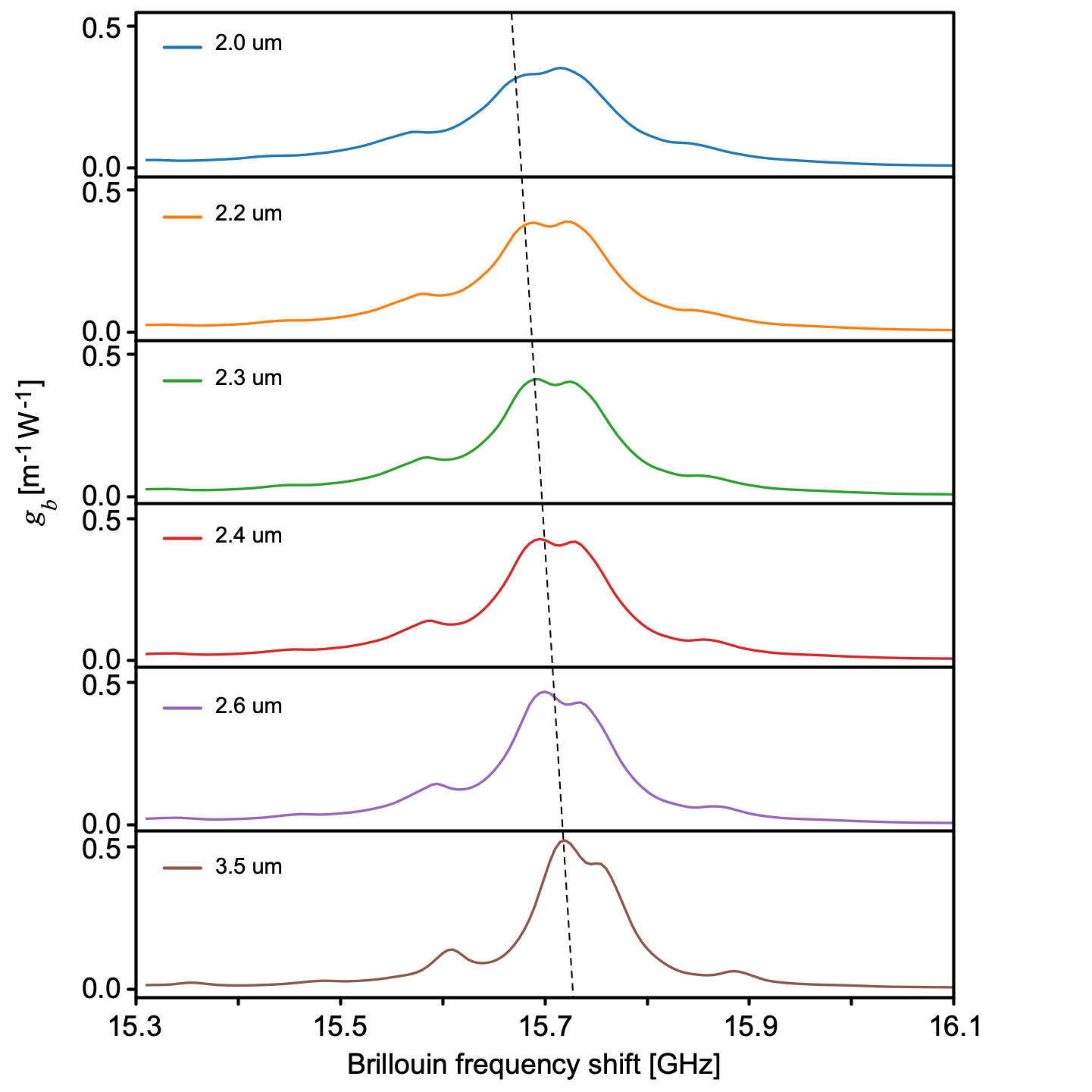}
\caption{Simulated SBS responses of the 2.0~µm, 2.2~µm, 2.3~µm, 2.4~µm, 2.6~µm, and 3.5~µm-wide SiON waveguides.}
\label{figs2}
\end{figure}

\subsection*{Comparison between the measurement and simulation results}
The simulated SBS responses of the 2.0~µm, 2.2~µm, 2.3~µm, 2.4~µm, 2.6~µm, and 3.5~µm-wide SiON waveguides are shown in Fig.~\ref{figs2}. The peak of the simulated SBS responses increases and shifts towards higher frequencies as the waveguide becomes wider, in the meantime, the linewidth also becomes narrower, all of which are coherent with the trend of our measurement results. 

We compared the Brillouin frequency shift, linewidth, and the Brillouin gain coefficient of different waveguides between the simulation and the measurement results in TABLE~\ref{tab:res}. The simulated Brillouin frequency shift is 1.3~GHz higher than the measured results, which is less than 10\% of the total frequency shift. The measured SBS linewidth of different waveguides are constantly larger than the simulations, however, the discrepancy keeps reducing as the waveguide becomes wider. The broader linewidth we measured could be contributed to the non-uniformity of the waveguides. There is also a discrepancy between the measured and the simulated Brillouin gain coefficients of different waveguides. The lower measurement values could come from the increasing coupling loss when we pump higher power into the samples. Nevertheless, the measured value also matches better with the simulation results for the wider waveguides. 

\begin{table}
\centering
\caption{\textbf{Simulated and measured Brillouin properties of SiON waveguides.}}
\label{tab:res}
\begin{tabular}{c|c|c|c|c|c|c}
    \textbf{Waveguide} & \multicolumn{2}{c|}{\textbf{Frequency shift}} & \multicolumn{2}{c|}{\textbf{Linewidth}} & \multicolumn{2}{c}{\textbf{Gain coefficient}} \\
    \textbf{width} & \textbf{Simulated} & \textbf{Measured} & \textbf{Simulated} & \textbf{Measured} & \textbf{Simulated} & \textbf{Measured}\\
    (µm) & (GHz) & (GHz) & (MHz) & (MHz) & (m$^{-1}$W$^{-1}$) & (m$^{-1}$W$^{-1}$)\\
    \hline
    2.0 & 15.70 & 14.22 & 154 & 358 & 0.35 & 0.10\\
    2.2 & 15.70 & 14.31 & 139 & 282 & 0.40 & 0.15\\
    2.3 & 15.71 & 14.40 & 131 & 280 & 0.42 & 0.17\\
    2.4 & 15.71 & 14.44 & 125 & 269 & 0.44 & 0.19\\
    2.6 & 15.71 & 14.43 & 115 & 213 & 0.47 & 0.24\\
    3.5 & 15.73 & 14.48 & 92 & 105 & 0.52 & 0.32
\end{tabular}
\end{table}

\begin{figure}
\centering
\includegraphics[width=0.8\linewidth]{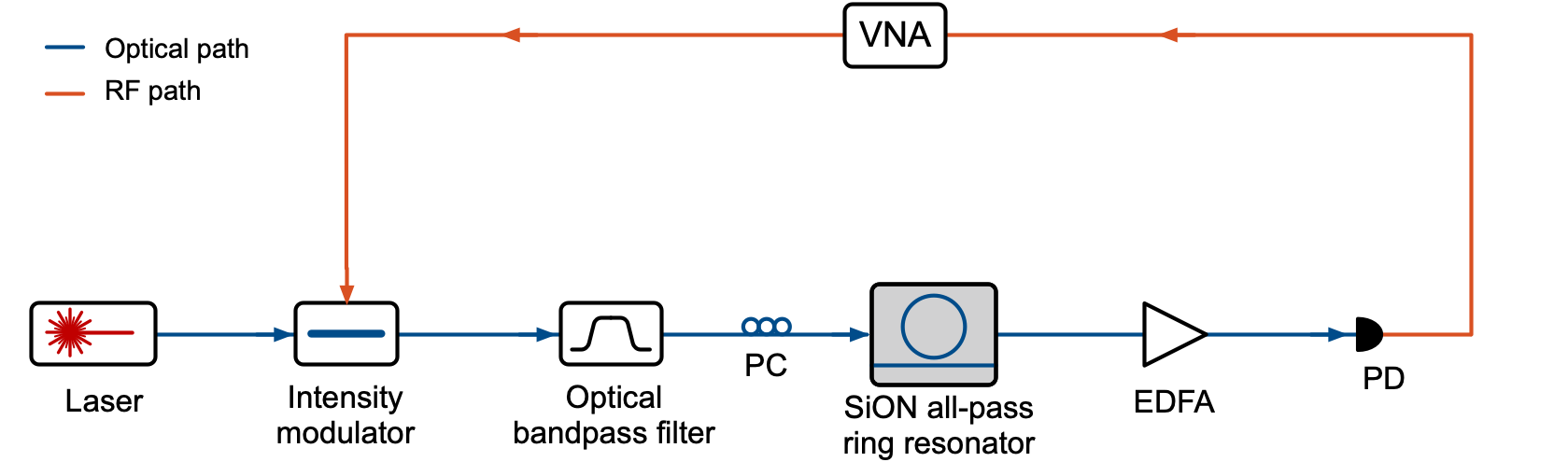}
\caption{Schematic of the setup used for ring resonator characterization. PC: polarization controller, PD: photodetector, EDFA: erbium-doped fiber amplifier, VNA: vector network analyzer.}
\label{figs3}
\end{figure}

\section*{Supplementary Note B: Details of the FWM Experiments}
\subsection*{Ring resonator characterization}
Before we performed the four-wave mixing (FWM) experiment, we first characterized the ring resonator with the experimental setup shown in Fig.~\ref{figs3}. The laser (Agere D2525P22) was modulated with an intensity modulator (Thorlabs LN05S-FC) and was sent to the optical bandpass filter (EXFO XTM-50) to filter out the lower sideband. After that, the light is coupled into the sample with the lensed fiber and coupled out from the other side of the chip. The signal was then amplified by the EDFA (Amonics AEDFA-33-B-FA) and converted to the RF domain with a photodiode (Optilab PD-23-C-DC). We swept the sideband of the light across the resonance response of the ring resonator using the vector network analyzer (VNA, Keysight P5007A) with an RF power of -5 dBm. From the measured S21 parameter, we then can get the linewidth of the ring resonator with MHz-level resolution, in addition, we can use phase information to confirm the ring is over-coupled.

\subsection*{FWM experiment setup}
\begin{figure}
\centering
\includegraphics[width=0.8\linewidth]{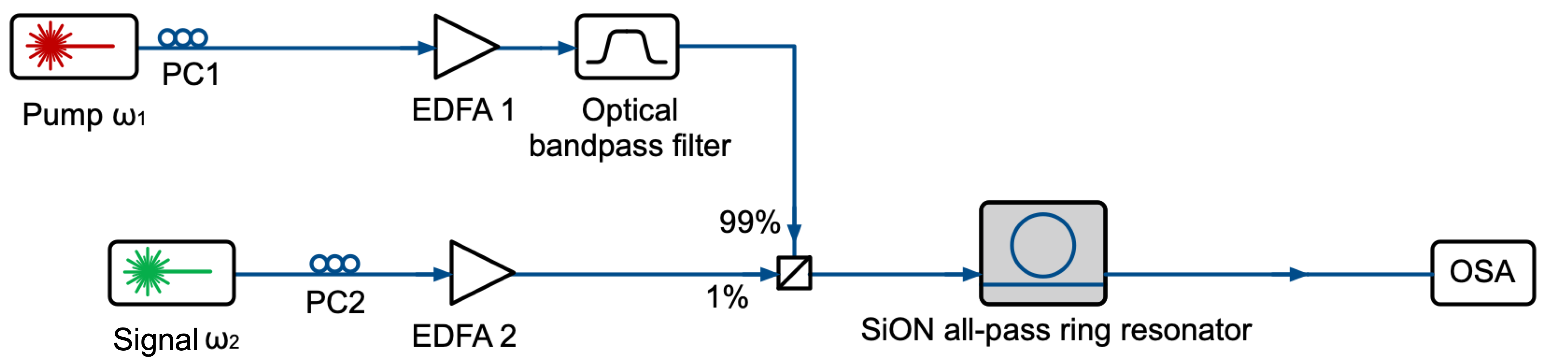}
\caption{Schematic of the setup used for four-wave mixing experiment. PC: polarization controller, EDFA: erbium-doped fiber amplifier, OSA: optical spectrum analyzer.}
\label{figs4}
\end{figure}

We measured the nonlinear index $n_2$ and nonlinear parameter $\gamma$ of the SiON waveguide with the FWM experiments in the all-pass ring resonator. The experimental setup was shown in Fig.\ref{figs4}. The pump laser (Santec TSL-210) operates at 1562~nm and is amplified with an EDFA (Amonics AEDFA-33-B-FA). After that, the pump is sent to an optical bandpass filter (EXFO XTM-50) to filter out the amplified spontaneous emission. The signal laser (Agere D2525P22) is set close to 2xFSR (100~GHz) away from the pump and is amplified with an EDFA (Amonics AEDFA-37-R-FA). The pump is thermally locked close to the resonance of the ring resonator, while the frequency of the probe is tuned manually. The pump and the signal are combined with a 99:1 optic coupler (Thorlabs TN1550R1A2) and coupled into the all-pass ring resonator with an AR-coated lensed fiber with a spot size of 2~$\mu$m (OZ optics). The generated idler together with the pump and signal is then coupled out from the chip and sent to the optical spectrum analyzer (Finisar Waveanalyzer 1500S) to measure the conversion efficiency from the signal to the idler. The experimental parameters are listed in TABLE~\ref{tab:FWM}.

\begin{table}
\centering
\caption{\textbf{The experimental parameters of the FWM characterization setup.}}
\label{tab:FWM}
\begin{tabular}{c|c|c|c}
    \textbf{Parameter} &\textbf{Value} & \textbf{Unit} &\textbf{Description}\\
    \hline
    $P_{\rm{pump}}$ & 20.3 & dBm & Input pump optical power \\
    $P_{\rm{signal}}$ & 6.3 & dBm & Input signal optical power \\
    $P_{\rm{ider}}$ & -38.7 & dBm & Output idler optical power \\
    $\delta v_{\rm{pump}}$ & 110 & MHz & Detuning of the pump light\\
    $\delta v_{\rm{signal}}$ & 197.5 & MHz & Detuning of the signal light\\
    $\delta v_{\rm{idler}}$ & 417.5 & MHz & Detuning of the idler light\\
    $FE_{\rm{pump}}$ & 31 & - & Enhancement factor of the pump light\\
    $FE_{\rm{signal}}$ & 25.2 & - & Enhancement factor of the signal light\\
    $FE_{\rm{ider}}$ & 12.4 & - & Enhancement factor of the idler light\\
    $L_{\rm{eff}}$ & 386 & mm & Effective length of the ring resonator
\end{tabular}
\end{table}

The conversion efficiency $\eta$ of the four-wave mixing in an all-pass ring resonator is \cite{Absil2000}:

\begin{equation}
\label{eq:eta}
\eta=\left(\gamma L_{\rm eff} P_p\right)^2\left|F E\left(\omega_p\right)\right|^4\left|F E\left(\omega_s\right)\right|^2\left|F E\left(\omega_i\right)\right|^2,
\end{equation}
where the $L$ is the circumference of the ring resonator, $P_p$ is the on-chip pump power, and the $FE(\omega_{p,s,i})$ is the field enhancement factor of the pump, signal, and idler, respectively.

The effective length considers both the attenuation and the phase mismatch of the four-wave mixing process:
\begin{equation}
\label{eq:el}
L^2_{\rm eff}=L^2 e^{-\alpha L}\left|\frac{1-\exp (-\alpha+i \Delta k L)}{\alpha L-i \Delta k L}\right|^2,
\end{equation}
where $\alpha$ is the propagation loss, and the $\Delta k$ is the phase mismatch between the pump, signal, and idler.

The field enhancement factor can be calculated by:

\begin{equation}
\label{eq:fe}
\left|FE\right|=\left|\frac{\sqrt{\kappa}}{1- \sqrt{1-\kappa} \exp (-\alpha L / 2) \cos \left(-\frac{2 \pi \delta v}{F S R}\right)}\right|,
\end{equation}
where the $\kappa$ is the power coupling coefficient and the $\delta v$ is the detuning. We calculated the detuning of the pump and the signal by comparing the extinction ratio of the light and the resonance response of the ring resonator. Assuming negligible dispersion, the detuning of the idler can be obtained based on:

\begin{equation}
\label{eq:detuning}
\delta {v_{\rm idler}} = 2 \delta {v_{\rm pump}} + \delta {v_{\rm signal}}
\end{equation}

The nonlinear parameter $\gamma$ can be calculated by combining (\ref{eq:eta} - \ref{eq:detuning}). Moreover, the nonlinear index $n_2$ can be calculated from the nonlinear parameter $\gamma$ by:
\begin{equation}
n_2 =\frac{c A_{\mathrm{eff}} \gamma}{\omega},
\end{equation}
where the $\omega$ is the angular frequency of the pump, $c$ is the speed of light in the vacuum and $A_{\rm eff}$ is the effective mode area of the waveguide.

\end{document}